\documentclass{aa}
\usepackage[varg]{txfonts}

\usepackage{graphicx}
\usepackage{dcolumn}
\usepackage{bm}

\usepackage{amsmath}
\usepackage{amssymb}

\usepackage{natbib}

\newcommand{\ep}{\epsilon}
\newcommand{\mm}{\mu_{max}}

\newcommand{\cvir}{c_{vir}}
\newcommand{\rvir}{R_{vir}}
\newcommand{\mvir}{M_{vir}}
\newcommand{\rc}{r_{conv}}

\begin{document}

\titlerunning{Relaxation of dark matter halos: how to match observational data?}
\title{Relaxation of dark matter halos: how to match observational data?}

\author{A. N. Baushev}
\offprints{baushev@gmail.com}
\author{Anton N. Baushev\inst{1, 2}}
\institute{DESY, 15738 Zeuthen, Germany\and
 Institut f\"ur Physik und Astronomie, Universit\"at Potsdam, 14476
Potsdam-Golm, Germany}

\date{}

\abstract{We show that moderate energy relaxation in the formation of dark matter halos invariably
leads to  profiles that match those observed in the central regions of galaxies. The density
profile of the central region is universal and insensitive to either the seed perturbation shape or
the details of the relaxation process. The profile has a central core; the multiplication of the
central density by the core radius is almost independent of the halo mass, in accordance with
observations. In the core area the density distribution behaves as an Einasto profile with low
index ($n\sim 0.5$); it has an extensive region with $\rho\propto r^{-2}$ at larger distances. This
is exactly the shape that observations suggest for the central region of galaxies. On the other
hand, this shape does not fit the galaxy cluster profiles. A possible explanation of this fact is
that the relaxation is violent in the case of galaxy clusters; however, it is not violent enough
when galaxies or smaller dark matter structures are considered. We discuss the reasons for this.}


\maketitle

\section{Introduction}
Although the dark matter component contributes most to the galaxy mass containment, a generally
accepted explanation for some observational properties of the galaxy dark matter halos has not been
supplied yet. In particular, there is some disagreement regarding the density profile in the center
of the halos. Earlier N-body simulation suggested very cuspy profiles with an infinite density in
the center (see, e.g., \citep{moore1999, neto2007}). Recent simulations \citep{mo09, navarro2010}
favor an Einasto profile with a finite central density, but the obtained Einasto index $n$ is so
high (typically $n\sim 5-6$) that the profile is very steep in the center and may still be called
cuspy. Although the simulations mainly model the largest structures of the Universe, such as galaxy
clusters ($M\sim 10^{15} M_\odot$), their results are expected to be valid for smaller objects as
well. Moreover, simulations of separate dark matter halos ($M\sim 10^{12} M_\odot$) have also be
performed (see, for instance, the {\it Via Lactea} project \citep{diemand2007}).

However, the correct interpretation of the N-body data requires a reliable estimation of the
simulation convergence. The idea of N-body simulations is to substitute real tiny dark matter
particles by heavy test bodies. Since there are fewer bodies, the task becomes computable. We face
a problem, however: the test bodies collide much more effectively than the original DM particles.
The strong encounters with high momentum transfer lead to evident effects such as kicking of the
test particles from the halo. They are avoided in simulations by softening of the Newtonian
potential near the test bodies. However, the gravitational force is long-acting, and the influence
of weak long-distant collisions dominates. In a nutshell, the gravitational potential of
homogeneous dark matter is plain, while the potential of the test bodies has local potential wells
near the bodies, despite of the softening. This produces unphysical soft scattering of the bodies
on each other and so leads to a collisional relaxation. The process can be described by the
Fokker-Planck approximation \citep{ll10}. The characteristic time of the collisional relaxation is
\citep[eqn. 1.32]{bt} $\tau_r= \dfrac{N(r)}{8 \ln\Lambda}\cdot\dfrac{r}{v}$, where $v$ is the
characteristic particle speed at radius $r$, $N(r)$ is the number of particles inside $r$,
$\ln\Lambda$ is the Coulomb logarithm. The ratio of the system lifetime $t_0$ to $\tau_r$ should be
low enough to guarantee the negligibility of the relaxation. A real halo contains $\sim 10^{65}$
particles, hence the collisions are wholly immaterial. The quantity of test bodies in simulations
is incomparably lower. The closer we approach the halo center, the smaller $N(r)$ and $r/v$ are,
and the shorter is $\tau_r$. Thus the central region of the halos is the most problematic for the
simulations: the profile inside some convergence radius $\rc$ may be already corrupted by the
collisions. It is commonly assumed that $\rc$ is defined by a certain value of $t_0/\tau_r$.

The commonly-used criterion of the convergence of N-body simulations is the stability of the
central density profile \citep{power2003}. The simulations indeed show that the central NFW-like
cusp is formed quite rapidly ($t<\tau_r$) and then is stable and insensitive to the simulation
parameters. However, the convergence criteria obtained with this method are surprisingly
optimistic: the cusp is stable at least up to $t=1.7 \tau_r$ and probably much longer
\citep{power2003}. \citet{hayashi2003} and \citet{klypin2013} reported that the cusp is stable even
at tens of relaxation times and smears out only at $t\sim 40 \tau_r$. The reasons why the collision
influence is negligible at a time interval exceeding the relaxation time are not quite clear.
Nevertheless, criterion $t=1.7 \tau_r$ \citep{power2003} is routinely used in modern simulations
\citep{navarro2010}.

The criteria based on the density profile stability have a weak point, however: the stability does
not guarantee the absence of the collisional influence. Considerations based on the Fokker-Planck
equation \citep{evans1997, 13} show that an NFW-like profile ($\rho\propto r^{-\beta}$,
$\beta\simeq 1$) is an attractor: the Fokker-Planck diffusion transforms any reasonable initial
distribution into it in a time shorter than $\tau_r$, and then the cuspy profile should survive
much longer than $\tau_r$, since the Fokker-Planck diffusion is self-compensated in this case.
Therefore the cusp is stable and insensitive to the simulation parameters; at $t\sim 50 \tau_r$ it
is destroyed by higher-order terms of the Boltzmann collision integral, disregarded by the
Fokker-Planck approach \citep{quinlan1996, 13}. This scenario perfectly describes the behavior of
real N-body simulations: the NFW-like cusp appears at $t<\tau_r$, remains stable up to tens of
relaxation times, and then is smoothed. However, the shape of the cusp is in this case defined by
the test particle collisions, that is, by a purely numerical effect. The only reliable criterion of
negligibility of the unphysical collisions is $t\ll\tau_r$. This means that $\rc$ is several times
larger than predicted by the criterion $t=1.7 \tau_r$ \citep{power2003}. Thus the criteria based on
the profile stability are most likely too optimistic and underestimate the influence of numerical
effects. The problem needs further investigation.

Contrary to the simulations, observations show a fairly smooth core in the centers of, at least,
galaxy halos \citep{deblok2001, bosma2002, marchesini2002, gentile2007}. \citet{mamon2011} removed
the baryon contribution and found that the dark matter distribution in the central regions of a
large array of galaxies may well be fitted by the Einasto profile with a low index ($n\simeq 0.5$)
that corresponds to a cored profile. The central densities of the dwarf spheroidal satellites of
the Andromeda galaxy are also low and favor the cored profiles \citep{tollerud2012}, although the
profiles in this case can be modified by the dynamical friction and tidal effects, since the
satellites are situated inside the virial radius of the host galaxy. However, recent observations
of dwarf spheroidal galaxies also indicate no cusps in their centers \citep{oh2011, governato2012}.
This makes attempts to explain the soft cores of the central density profiles by the influence of
the baryonic component dubious: the dwarfs contain only a very minor fraction of baryons.

Many galaxies (at least, the spiral ones) show quite an extensive region in their dark matter halo
with a $\rho\propto r^{-2}$ profile: the region corresponds to a characteristic flat tail in their
rotation curves. This feature allowed proving the existence of the dark matter by \citet{rubinold}.
Meanwhile, none of two profiles (Navarro-Frenk-White or Einasto) that are commonly used to fit the
halos in the N-body simulations has such a region. Certainly, the current power-law index $\gamma=
d\log\rho/d\log r$ of both the profiles reaches $-2$ at some point. However, the index changes
continuously in both cases, the point where $\gamma=-2$ is marked not, therefore we cannot expect
an extensive region with $\rho\sim r^{-2}$. Of course, the real structure formation is a much more
complex process than the simulations, and the origin of the region could be a result of the
influence of the baryon component, substructures, galaxy disk, etc. However, the persistence of the
isothermal-like shape $\rho\sim r^{-2}$ in the density profiles of a vast collection of galaxies
with very different physical properties \citep{rubinnew} suggests a more fundamental and more
universal physical reason.

Finally, observations indicate that the multiplication of the halo central density $\rho_c$ by the
core radius $r_{core}$ is almost constant for a wide variety of galaxies, while their physical
parameters, including $\rho_c$ and $r_{core}$ apart, change in a rather extensive range. This
effect was first discovered by \citet{kormendy2004} and then confirmed by several independent
observations (see \citet{salucci2007, salucci2009} and references therein). To be able to compare
results obtained using different profile models, we define the core radius $r_{core}$ as the
radius, at which
 \begin{equation}
 \label{16a0}
\frac{d\log\rho(r_{core})}{d\log r}=-1.
\end{equation}
\citet{salucci2009} used the Burkert profile \citep{burkert}
 \begin{equation}
 \label{16a1}
\rho(r) = \dfrac{\rho_c r^3_b}{(r+r_b)(r^2+r^2_b)}.
\end{equation}
It is easy to see that $r_{core}=r_b/2$.  Recently, \citet{salucci2009} found that $\log(\rho_c
r_b)=2.15\pm 0.2$ in units of $\log (M_\odot \text{pc}^{-2})$ on the basis of the co-added rotation
curves of $\sim 1000$ spiral galaxies, the mass models of individual dwarf irregular and spiral
galaxies of late and early types with high-quality mass profiles, and the galaxy-galaxy
weak-lensing signals from a sample of spiral and elliptical galaxies. They also showed that the
observed kinematics of Local Group dwarf spheroidal galaxies are consistent with this value as
well. The result was obtained for galactic systems belonging to various Hubble types whose mass
profiles have been determined by several independent methods.

The aim of this article is to show that all the above-mentioned features (a cored central profile,
an extended region with a $\rho\propto r^{-2}$ profile,  and $\rho_c r_{core}\simeq\it{const}$
relationship) appear automatically, if we assume that the relaxation of the galactic halos during
their formation was not violent. The violent-relaxation scenario, usually leading to a cuspy
density profile, was first suggested by \citet{violent} for stellar systems. The idea of it is that
strong small-scale gravitational fields appear during the halo relaxation, and as a result all the
particles completely forget their initial states. Recent N-body simulations \citep{diemand2005,
diemand2007, diemandkuhlen2008} showed however, that this assumption is probably incorrect, and a
significant part of the particles and subhalos 'remember' their initial specific energies
$\ep=\frac{v^2}{2}+\phi$: they change quite moderately.

There may be several theoretical reasons for the absence of the violent relaxation \citep{15}. For
instance, the efficiency of the violent relaxation rapidly drops with the growth of the initial
radius $r$ of the area under consideration from the center of the object. Even the original paper
\citep{violent} reported that the outer regions of the stellar clusters remained unrelaxed.
Meanwhile, a dark matter halo originates from a perturbation that was initially linear and, in
contrast to the formed structures, had a low density contrast. Consequently, the main contribution
to the halo mass was made by the layers with large $r$, since their volume $4 \pi r^2 dr$
dominates. This circumstance impedes the relaxation. Moreover, a significant part of the dark
matter gradually accretes onto the already formed halo, when the strong gravitational field
inhomogeneities have already disappeared \citep{wang2012}.

All these reasons allow us to assume that the violent relaxation does not occur, at least, in some
types of halos.  Hereafter we assume that the relaxation of \emph{low-mass} halos ($M_{vir}\lesssim
10^{12} M_\odot$), corresponding to galaxies, is not violent. We assume that the relaxation is
moderate in the following sense:
\begin{enumerate}
 \item \emph{The final total specific energy $\ep_f$ of most of the particles differs from
the initial ones $\ep_i$ no more than by a factor $\cvir/5$}
 \begin{equation}
 \label{16b1}
\frac{\ep_f}{\ep_i}\le\frac{\cvir}{5}.
\end{equation}
\item \emph{There can be particles that violate condition (\ref{16b1}), but their total mass should
be small with respect to the halo mass inside $r=\frac{2\rvir}{\cvir}$
\begin{equation}
 \label{16b2}
M<\int\limits_0^{2\rvir/\cvir} dM_{\text{halo}}.
\end{equation}
The reason for this limitation will be clear from the subsequent text.}
\end{enumerate}

Here we used the NFW halo concentration $\cvir$. As we will see, the real density profile may
significantly differ from the NFW one, if conditions (\ref{16b1})-(\ref{16b2}) are true. However,
we use $\cvir$ because of its popularity and in view of the fact that characteristic values of
$\cvir$ for various types of astronomical objects are well known. Condition (\ref{16b1}) is too
strict for galaxy clusters, since their concentrations are low ($\cvir\sim 3-5$). Indeed, even for
rather a dense cluster ($\cvir =6$) Eq.~(\ref{16b1}) would mean that the energies of almost all the
particles change by no more than $20\%$ during the relaxation. The real relaxation is most likely
more intensive; perhaps, this is the reason why the galaxy clusters have profiles close to NFW
\citep{okabe2010}.

In contrast, conditions (\ref{16b1}) and (\ref{16b2}) seem quite soft for galactic halos.
Concentration $\cvir\simeq 12-17$ even for the giant Milky Way galaxy and probably much higher for
low-mass galaxies. Consequently, assumption (\ref{16b1}) means that the energy of most of the
particles changes no more than by a factor of $3$ with respect to the initial value. This behavior
looks quite natural for a collisionless system. Condition (\ref{16b2}) is also weak. Indeed, a halo
of typical galactic concentration ($\cvir\sim 15-20$) contains $20-25\%$ of its mass inside
$r=\dfrac{2\rvir}{\cvir}$; this means that condition (\ref{16b2}) reduces to the constraint that
the fraction of the particles that changed their energy by more than a factor of $\cvir/3\simeq 3$
is smaller than a quarter of the halo mass.

Of course, there is always some dark matter that violates condition (\ref{16b1}). For instance, the
particles that were in the center of the halo at the very beginning of the collapse, when their
velocities (as well as the velocities of other particles) were low. Their energies changed by much
more than Eq.~(\ref{16b1}) during the collapse, even if there was no relaxation at all. Indeed,
they remain in the halo center during the collapse, while the gravitational potential of this area
deepens by approximately a factor $\cvir$ because of the crowding of the matter toward the center.
However, the density of these 'ancient habitants' of the halo center was comparable with the
average DM density of the Universe at the moment of the halo collapse, that is, it was only by a
factor $\sim 5$ higher than the present-day value. Meanwhile, the central density of the Milky Way
is higher by a factor $\sim 3\cdot 10^5$ than the Universe DM density. Clearly,  the 'ancient
habitants' do yield some density into the DM content of the Galactic center, but the contribution
is negligible ($\sim 10^{-5}$).

To conclude the introductory section, we should emphasize that the moderate relaxation is now no
more than a hypothesis. However, as we will see, it leads to quite correct predictions of the
central density profiles of galaxies.

In Sect.~\ref{16sec2} we discuss the energy evolution of a collapsing dark matter halo and show
that the distribution of the formed halo probably has a peculiar form. In Sect.~\ref{16sec3} we
calculate the density profile corresponding to this distribution. In Sect.~\ref{16sec4} we discuss
the obtained profile and compare it with observational data.  Finally, in Sect.~\ref{16sec4}, we
briefly summarize our results and discuss further implications.

\begin{figure}
 \resizebox{\hsize}{!}{\includegraphics[angle=0]{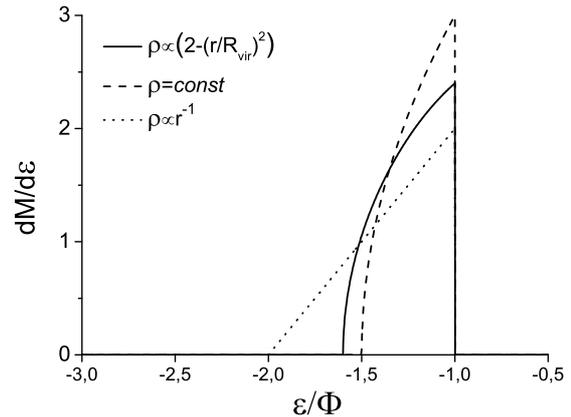}}
\caption{Initial energy spectra $\frac{dM}{d\ep}$ for three shapes of initial perturbation:
$\rho\propto \left(2-\left(\frac{r}{\rvir}\right)^2\right)$ (solid line), $\rho=\text{\it const}$
(dashed line), $\rho\propto r^{-1}$ (dotted line). In all the cases the spectra are similar,
narrow, and strongly concentrated toward $\ep=-\Phi$.}
 \label{fig1}
\end{figure}

\section{Energy distribution}
\label{16sec2} The assumption of the moderate energy evolution immediately leads to some important
consequences. Hereafter we accept for simplicity that the halo is spherically symmetric. The
present-day dark matter halos were formed from some primordial perturbations that existed in the
early Universe. Initially, the perturbations grew linearly, but then they reached the nonlinear
regime and collapsed. We consider the initial (at the moment when $\delta \rho/\rho\simeq 1$)
energy distribution of the particles that later formed the halo. It is determined by the
gravitational field of the initial perturbation, and the closer a particle was to the center, the
lower was its energy. However, the potential well of the initial perturbation cannot be deep, and
the particles cannot have a very low energy, because thy are lumped in quite a narrow energy
interval. Indeed, a trivial Newtonian calculation gives us the initial energy spectrum for various
shapes of initial perturbations \citep{15}. We assume for simplicity that the size of the
perturbation at the moment of the collapse is equal to the virial radius of the formed halo
$\rvir$: these values should be similar in the very general case \citep{gorbrub2}. We introduce the
virial potential of the halo $\Phi=G\dfrac{\mvir}{\rvir}$. As an example, we consider the case when
the initial density distribution of the perturbation has the shape $\rho\sim r^{-1}$ inside
$\rvir$. Then $M(r)=\int 4 \pi r^2 \rho dr=\mvir (r/\rvir)^2$ and
\begin{equation}
 \label{16b3}
\dfrac{d\phi(r)}{dr}=G \dfrac{M(r)}{r^2}= G \dfrac{\mvir}{\rvir^2}\qquad \phi(r)= - \Phi\left(2-
\dfrac{r}{\rvir}\right).
\end{equation}
Here we took into account that $\phi(\rvir)=-\Phi$. It follows from the general cosmological
consideration that the initial velocity of the matter may be thought to be zero without loss of
generality \citep{gorbrub2}. Therefore, the specific total energy of a particle is equal to the
specific gravitational energy, that is, to the gravitational potential $\ep=\phi(r)$. By dividing
$dM(r)=\dfrac{2r \mvir}{\rvir^2} dr$ by $d\ep=d\phi(r)=G \dfrac{\mvir}{\rvir^2}dr$, we obtain
\begin{equation}
 \label{16b4}
\dfrac{dM}{d\ep}=\dfrac{dM}{d\phi(r)}=\dfrac{2r}{G}=\frac{2\mvir}{\Phi}
\left(2+\frac{\ep}{\Phi}\right).
\end{equation}
In a similar manner, we can obtain the initial energy spectra for various forms of initial
perturbations (see \citep{15} for details). Distributions $\frac{dM}{d\ep}$ for three different
shapes ($\rho\propto \left(2-\left(\frac{r}{\rvir}\right)^2\right)$ (solid line), $\rho=\text{\it
const}$ (dash line), $\rho\propto r^{-1}$ (dotted line)) are represented in Fig.~\ref{fig1}. In all
the cases the spectra are quite similar, narrow, and strongly concentrated toward $\ep=-\Phi$. Even
an unphysically steep initial perturbation $\rho\sim r^{-1}$ contains only particles with
$\ep\in[-2\Phi;-\Phi]$.

We now consider the formed halo. Its gravitational field is stationary.  The motion of a particle
in the central gravitational field $\phi(r)$ can be explicitly characterized by two integrals of
motion: its specific angular momentum $\mu\equiv |[\vec v\times\vec r]|$ and specific energy
$\ep=\dfrac{v_r^2}{2}+\dfrac{\mu^2}{2 r^2}+\phi$. Instead of $\ep$, it is more convenient to use
the apocenter distance of the particle $r_0$, which is the largest distance that the particle can
move away from the center. It is bound with $\ep$ as $\ep=\phi(r_0)+\mu^2/(2 r_0^2)$. We may
introduce distribution function $f(r_0)$ of the particles of the formed halo over $r_0$
\begin{equation}
 \label{16a2}
dm=f(r_0) dr_0\qquad \int^{R_{vir}}_0 f(r_0) dr_0 = M_{vir}.
\end{equation}
As we will see, if conditions (\ref{16b1})-(\ref{16b2}) are satisfied, $f(r_0)$ has a very peculiar
appearance.

It is extremely important for our consideration that the potential well of the collapsed halo is
much deeper then the initial one. The depth (i.e., the value of $\phi(0)$) depends on the halo
profile: for the NFW $\phi(r)\simeq -\cvir\Phi$, if $\cvir\gg 1$ \citep{clump}. Although density
profiles of real galaxies are much more complex, $\phi(r)\simeq -\cvir\Phi$ may still be a good
approximation. For instance, we may accept for the Milky Way $\mvir=10^{12} M_\odot$,
$\rvir=250$~{kpc}, $\cvir\simeq 15$ \citep{klypin}: then $\Phi\simeq (130 \text{km/s})^2$.
Meanwhile, the Galaxy escape speed near the solar system unambiguously exceeds $525$~{km/s}
\citep{escape} and may in principle be much higher ($650$~{km/s} or even higher \citep{suchkov,
bt}). Accordingly, $|\phi(0)|\gtrapprox \cvir |\phi(\rvir)|=\cvir \Phi$.

As we showed, initial energy spectra are very similar for any reasonable shape of the initial
perturbation. We consider for the sake of definiteness an initial perturbation $\rho\propto
\left(2-\left(\frac{r}{\rvir}\right)^2\right)$, because it seems to be a good approximation for a
real one. We can see in Fig.~\ref{fig1} that most the particles have $\ep\simeq -\Phi$, and there
are no particles with $\ep < -1.6\Phi$. Consequently, the particles obeying condition (\ref{16b1})
may not have $\ep < -\dfrac{\cvir}{3}1.6\Phi\simeq -\dfrac{\cvir}{2}\Phi$, and even the fraction of
the particles with $\ep\simeq -\dfrac{\cvir}{2}\Phi$ is small: the initial spectrum contains only a
few particles with $\ep\simeq -1.6\Phi$. We estimate $r_0$, corresponding to a particle with $\ep=
-1.6\Phi$. Of course, it depends on the density profile of the halo and on the particle angular
momentum. The influence of the latter factor can be easily taken into account: a nonzero angular
momentum decreases $r_0$ of a particle of a given energy $\ep$, but cannot decrease it by a factor
exceeding $2$. Indeed, a particle of given energy $\ep$ in a given central gravitational field has
the largest $r_0$ if $\mu=0$, and its orbit is radial. The ratio of $r_0$ for the radial and the
circular cases is $2$ in the instance of the gravitational field of a point mass. It is easy to see
that the ratio can only be lower, if we consider a distributed density profile instead of a point
mass: if the dark matter is spread, the more compact circular orbit encloses a smaller central mass
than in the point-mass case. Moreover, N-body simulations suggest that the orbits of most of the
particles are elongated \citep{2006JCAP...01..014H}. In this case the influence of the angular
momentum on $r_0$ is negligible.

A particle of energy $\ep\simeq -\dfrac{\cvir}{2}\Phi$ has $r_0\simeq \dfrac{2\rvir}{\cvir}$ in the
case of an NFW profile. The potential wells of real galaxies are most likely deeper than the
best-fit NFW predicts (probably because of the influence of the much more concentrated baryon
component). For instance, the escape velocity from the center of an NFW halo with $\mvir=10^{12}
M_\odot$ and $\cvir\simeq 15$ (as we could see, these values approximately correspond to the dark
matter halo of the Milky Way) is $\simeq 300$~{km/s}, while the real escape speed from the center
of the Galaxy is at least twice as high \citep{escape}. The deeper the potential well, the larger
is $r_0$ that corresponds to the same $\ep$; consequently, $r_0= \dfrac{2\rvir}{\cvir}$ is a
conservative estimate of $r_0$ of a particle with energy $\ep\simeq -\dfrac{\cvir}{2}\Phi$.
Consequently, particles obeying condition (\ref{16b1}) cannot have smaller $r_0$. However, the halo
contains a significant part of its mass ($\sim 25\%$) inside $r = \dfrac{2\rvir}{\cvir}$.

The absence of violent relaxation leads to a very important consequence: the density profile in the
center of the halo is formed by the particles that arrive from the outside. Since their $r_0$ are
larger than $\dfrac{2\rvir}{\cvir}$, some part of their trajectories lie outside of this area. Of
course, the real situation is more complex, and there are always particles that violate condition
(\ref{16b1}). However, condition (\ref{16b2}) guarantees that their contribution inside $r =
\dfrac{2\rvir}{\cvir}$ is small. As we could see, condition (\ref{16b2}) is quite soft for the real
systems.

We can also expect that most of the particles have $r_0\sim\rvir$. Indeed, the initial energies of
the particles were very close to the virial one $\Phi=-G\frac{\mvir}{\rvir}$. Since the total
energy of the system is conserved, the average energy of the particles remain close to $-\Phi$;
consequently, all the particles may not drop their energies by a factor $\sim \cvir/5$. The
particle energy exchange is a more or less random process, and we may expect that the particle
energy near the average value $-\Phi$ is much more probable than the minimum possible
$-\dfrac{\cvir}{5} \Phi$. Consequently, even the fraction of particles with $r_0\simeq
\dfrac{2\rvir}{\cvir}$ should be small. However, this statement is less rigid (and less important
for us) than the above-mentioned dominance of the particles with $r_0\gg r$ in the center.

A region with a dominant fraction of particles with $r_0\gg r$ inevitably occurs in the center of
the halo, if the relaxation is moderate. It even appears if the relaxation is much more violent
than Eq.~(\ref{16b1}) (for instance, if $\ep_f/\ep_i\sim \cvir/2$): the lowest energy of the
particles is still higher than $\phi(0)\simeq -\cvir\Phi$ in this case. However, the radius of the
area is then much smaller. Hereafter we use condition (\ref{16b1}), and the area is quite large in
this case: $\dfrac{2\rvir}{\cvir}\simeq 30$~{kpc} for the Milky Way galaxy.

\section{Calculations}
\label{16sec3} The density distribution in the center of the halo, created by the particles that
arrive in this region from the outside, is universal and insensitive to the shape of distribution
$f(r_0)$ \citep{15, 14}. First of all, we specify the angular momentum distribution of the
particles. According to results of the numerical simulations (see, for instance,
\citep{kuhlen2010}), the distribution over $v_\tau$ deviates, but is still similar to Gaussian. We
assume for simplicity that their specific angular momentum has a Gaussian distribution
\begin{equation}
 \label{16a3}
dm \propto \frac{2\mu}{\alpha^2} \exp\left(-\frac{\mu^2}{\alpha^2}\right)\; d\mu,
\end{equation}
where $\alpha\equiv \alpha(r_0)$ is the width of the distribution, depending on $r_0$. The total
distribution can be rewritten as
\begin{equation}
 \label{16a4}
dm = f(r_0) \frac{2\mu}{\alpha^2} \exp\left(-\frac{\mu^2}{\alpha^2}\right)\; dr_0 d\mu.
\end{equation}
As we have already mentioned, most of the particles have $r_0\sim R_{vir}$, and therefore only
those with small $\mu$ can penetrate into the area of our interest $r\sim r_{core} \ll R_{vir}$.
This means that the distribution in the halo center is mainly determined by the behavior of
Eq.~(\ref{16a4}) at $\mu\simeq 0$, where Eq.~(\ref{16a4}) is finite. Thus our calculation is not
sensitive to the distribution over $\mu$: we would obtain a very similar result for any other
distribution, which has the same value $2\mu f(r_0)/\alpha^2(r_0)$ when $\mu \to 0$.

Since the particle energy conserves
 \begin{equation}
 \label{16a5}
\dfrac{\mu^2}{2 r_0^2}+\phi(r_0)=\ep=\dfrac{v_r^2}{2}+\dfrac{\mu^2}{2 r^2}+\phi(r),
\end{equation}
the radial and tangential components of the particle velocity are equal to
 \begin{equation}
 \label{16a6}
 v_r = \sqrt{2(\phi(r_0)-\phi(r))-\mu^2\left(\frac{1}{r^2}-\frac{1}{r_0^2}\right)}\qquad v_\tau=
 \frac{\mu}{r}.
\end{equation}
The zero of the radicand gives us the maximum angular momentum of a particle with which it can
reach radius $r$,
 \begin{equation}
 \label{16a7}
 \mm^2 = 2 (\phi(r_0)-\phi(r)) \left(\frac{1}{r^2}-\frac{1}{r_0^2}\right)^{-1}.
\end{equation}
We may rewrite Eq.~(\ref{16a6}) as
 \begin{equation}
 \label{16a8}
 v_r = \frac{\sqrt{r_0^2-r^2}}{r r_0} \:\sqrt{\mm^2-\mu^2}.
\end{equation}
We also need the half-period of the particle, that is, the time it takes for the particle to fall
from its largest to the smallest radius,
 \begin{equation}
 \label{16a9}
 T(r_0,\mu) = \int^{r_0}_{r_{min}}\;\frac{dr}{v_r}.
\end{equation}
$T$ is, generally speaking, a function of $r_0$ and $\mu$. However, as was shown in \citet{15}, for
the particles that can reach the cental region the dependence on $\mu$ is extremely weak (the
reason is that the function $T(r_0,\mu)$ slowly changes near the extremum at $\mu=0$). Therefore we
may approximate $T(r_0,\mu)\simeq T(r_0,0)\equiv T(r_0)$.

A particle of mass $m$ contributes to the halo density throughout the interval between $r_0$ and
the smallest radius the particle can reach. The contribution in an interval $dr$ is proportional to
the time the particle spends in this interval \citep{11}
 \begin{equation}
 \label{16a10}
 \frac{dm}{m} =  \frac{\delta\rho \cdot 4\pi r^2 dr}{m}=\frac{dt}{T(r_0)} = \frac{dr}{v_r T(r_0)},
\end{equation}
Here $\delta\rho$ is the contribution of the particle to the total halo density at radius $r$. We
obtain that $\delta\rho=\frac{m}{4\pi r^2 v_r T(r_0)}$. To determine the total halo density, we
substitute here a mass element (\ref{16a4}) instead of $m$ and integrate over $dr_0$ and $d\mu$,
 \begin{equation}
 \label{16a11}
\rho = \int\limits^{R_{vir}}_0 \frac{f(r_0) dr_0}{4 \pi r^2 T(r_0) \alpha^2(r_0)}
\int\limits^{\mm}_0\!\frac{2 \mu}{v_r} \exp\left(-\dfrac{\mu^2}{\alpha^2(r_0)}\right)\; d\mu.
\end{equation}
If we substitute equation (\ref{16a8}) for $v_r$, the second integral can be taken analytically:
 \begin{equation}
 \label{16a12}
\int\limits^{\mm}_0\!\frac{2 \mu}{v_r}\exp\left(-\dfrac{\mu^2}{\alpha^2(r_0)}\right)\; d\mu =
\frac{2 r r_0 \alpha(r_0)}{\sqrt{r_0^2-r^2}} \; D\left(\dfrac{\mm}{\alpha(r_0)}\right),
\end{equation}
where $D(x)\equiv e^{-x^2}\int_0^x e^{t^2} dt$ is the Dawson function. Since we calculate the
density profile of the central region and use assumption $r\ll r_0$, we can significantly simplify
Eq.~(\ref{16a12}). In particular, it follows from Eq.~(\ref{16a7}) that $\mm\simeq r \sqrt{2
(\phi(r_0)-\phi(0))}$. We obtain
 \begin{equation}
 \label{16a13}
\rho = \int^\infty_0\!\frac{f(r_0)}{2 \pi \alpha(r_0) T(r_0) r}\; D\left(r\dfrac{\sqrt{2
(\phi(r_0)-\phi(0))}}{\alpha(r_0)}\right) dr_0.
\end{equation}
We can factor out the Dawson function from the integral using the above-mentioned properties of
function $f(r_0)$ (see the end of the {\it Energy distribution} section). First, it is almost equal
to zero for small $r_0$: this means that the integration in Eq.~(\ref{16a13}) is performed not from
$0$, but from $\dfrac{2\rvir}{\cvir}$. Second, as we could see, the formed halo is dominated by the
particles with $r_0\sim R_{vir}$. It follows that the main contribution to the integral in
Eq.~(\ref{16a13}) is given by the part close to the upper limit $r_0\simeq R_{vir}$: roughly
speaking, by $r_0\in [R_{vir}/2;R_{vir}]$. These two properties of $f(r_0)$ mean that $f(r_0)$
sharply depends on $r_0$ at this interval. Conversely, $\alpha(r_0)$ probably does not change much
in interval $[R_{vir}/2; R_{vir}]$: $\alpha(r_0)$ is widely believed to be a power-law dependence
with the index between $-1$ and $1$ \citep{2006JCAP...01..014H}. $\sqrt{2 (\phi(r_0)-\phi(0))}$
changes even more slowly: for instance,
$\sqrt{(\phi(R_{vir})-\phi(0))/(\phi(R_{vir}/2)-\phi(0))}\simeq 1.13$ for the NFW profile with
$c_{vir}=15$. Moreover, $D$ is a finite and not very sharp function of its argument. Comparing this
with the sharp behavior of $f(r_0)$, we may neglect the weak dependence of the argument of function
$D$ in Eq.~(\ref{16a13}) on $r_0$ and substitute some value, averaged over the halo (see the
$Appendix$ for details),
 \begin{equation}
 \label{16a14}
r_c = \left\langle\frac{\alpha(r_0)}{\sqrt{2
(\phi(r_0)-\phi(0))}}\right\rangle\simeq\frac{\langle\alpha(r_0)\rangle}{\sqrt{2 |\phi(0)|}}.
\end{equation}
Then we can rewrite Eq.~(\ref{16a13}) and obtain the final result:
 \begin{equation}
 \label{16a15}
\rho = \rho_c \frac{r_c}{r}\: D\!\left(\frac{r}{r_c}\right), \qquad \rho_c=\frac{1}{2 \pi
r_c}\int^\infty_0\!\frac{f(r_0) dr_0}{\alpha(r_0) T(r_0)}.
\end{equation}
Since $D(r/r_c)\simeq r/r_c$, when $r/r_c \to 0$, $\rho_c$ is really the central density of the
halo. As we can see, it is always finite. At the same time, the shape of the density profile only
depends on parameter $r_c$.

 \begin{figure}
 \resizebox{\hsize}{!}{\includegraphics[angle=0]{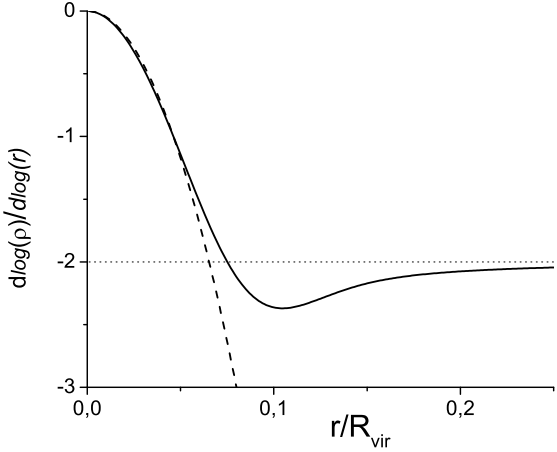}}
\caption{Density profile of the model under consideration (\ref{16a15}) (with $r_c=0.05 R_{vir}$,
solid line). An Einasto profile with $n=0.5$ and $r_s=0.017 R_{vir}$ is plotted for comparison
(dashed line).}
 \label{16fig2}
\end{figure}

\section{Discussion}
\label{16sec4} Fig.~\ref{16fig2} represents profile (\ref{16a15}) with $r_c=0.05 R_{vir}$ (solid
line). An Einasto profile with $n=0.5$ and $r_s=0.017 R_{vir}$ is plotted for comparison (dashed
line). The model profile is very similar to the Einasto profile in the center; consequently, it
fits experimental data well \citep{mamon2011}. The second consequence is that Eq.~(\ref{16a15})
describes a cored profile with $r_{core}\simeq r_c$; using criterion (\ref{16a0}), we obtain
$r_{core}\simeq 0.924 r_c$.

Profile (\ref{16a15}) in all conditions transforms into $\rho\propto r^{-2}$ at large distances,
which may explain the persistence of the flat regions in the rotation curves of a vast collection
of galaxies with very different physical properties. However, a question arises: we assume that
$r_0\gg r$ during the derivation of Eq.~(\ref{16a15}). Is this approximation (and, consequently,
Eq.~(\ref{16a15})) still valid for large $r$, where the profile behaves as $\rho\propto r^{-2}$? As
we showed, Eq.~(\ref{16a15}) is valid for $r<\dfrac{2\rvir}{\cvir}$. Meanwhile, the $\rho\propto
r^{-2}$ profile appears if $r\gg r_c$. The value of $r_c$ depends on $\langle\alpha\rangle$ and
$\sqrt{2 |\phi(0)|}$ according to Eq.~(\ref{16a14}). We show below that the best agreement between
the theory and observations is achieved if $\langle\alpha\rangle$ is respectively small
(\ref{16a27}). Substituting Eq.~(\ref{16a27}) and $|\phi(0)|\simeq \cvir\Phi$ into (\ref{16a14}),
we can roughly estimate $r_c\simeq\rvir/(28\sqrt{2\cvir})$. We illustrate this on the example of
the Milky Way galaxy. Here $\dfrac{2\rvir}{\cvir}\simeq 30$~{kpc}, that is, approximately the disk
radius; $r_c\simeq\rvir/(28\sqrt{2\cvir})\simeq 1.6$~{kpc}, which is comparable with the bulge
size. So $\dfrac{2\rvir}{\cvir}\gg r_c$, equation (\ref{16a15}) is still valid for $r\gg r_c$, and
we may expect an extended region with $\rho\propto r^{-2}$ between $\sim 2 r_c \simeq 3.2$ and
$\sim 30$~{kpc}.

Now we can investigate how the multiplication $\rho_c r_c$ depends on the halo mass in our model.
According to Eq.~(\ref{16a15}),
 \begin{equation}
 \label{16a16}
 \rho_c r_c=\frac{1}{2 \pi}\int^\infty_0\!\frac{f(r_0) dr_0}{\alpha(r_0) T(r_0)}.
\end{equation}
We can significantly simplify this equation with the help of the same technic that we used to
transform equation (\ref{16a13}) into (\ref{16a15}) (see Appendix): neglect the fairly weak
dependencies of functions $\alpha(r_0)$ and $T(r_0)$ on $r_0$ (compared with $f(r_0)$), and
substitute some values, averaged over the halo. Then equation (\ref{16a15}) may be rewritten as
 \begin{equation}
 \label{16a17}
\rho_c r_c \simeq \frac{1}{2\pi}\frac{\int^\infty_0\!f(r_0) dr_0}{\langle\alpha\rangle\langle
T\rangle}=\frac{1}{2\pi}\frac{M_{vir}}{\langle\alpha\rangle\langle T\rangle}.
\end{equation}

Now we estimate $\langle\alpha\rangle$ and $\langle T\rangle$. To begin with, we assume that
$\langle\alpha\rangle$ has the highest possible value: it can hardly be higher than
\begin{equation}
 \label{16a18}
 \langle\alpha\rangle=\frac14 \sqrt{G M_{vir} R_{vir}}
\end{equation}
because a significant fraction of the halo particles would not be gravitationally bound in the
opposite case. Below we discuss the applicability of this assumption.

The half-period $T(r_0)$ is mainly determined by the gravitational acceleration at $r_0$ (where
$v_r=0$) and is not very sensitive to the density distribution in the halo center. As we showed, a
significant part of the particles should have $r_0\sim\rvir$. Therefore we accept as an estimate of
$\langle T\rangle$ the time necessary for a particle with no angular momentum to fall from
$r=\rvir/2$ on a point mass,
\begin{equation}
 \label{16a19}
 \langle T\rangle=\int^{\rvir}_0\frac{dr}{v_r}=\dfrac{\pi}{8} \dfrac{\rvir^{3/2}}{\sqrt{G \mvir}}.
\end{equation}
Indeed, $\langle r_0\rangle$ can hardly be smaller than $\rvir/2$: as we showed, $r_0\sim \rvir$
for most of the particles in our model. On the other hand, we underestimate $\langle T\rangle$ by
considering a point mass instead of the real distribution. Consequently, equation (\ref{16a19})
most likely underestimates $\langle T\rangle$.

\begin{figure}
 \resizebox{\hsize}{!}{\includegraphics[angle=0]{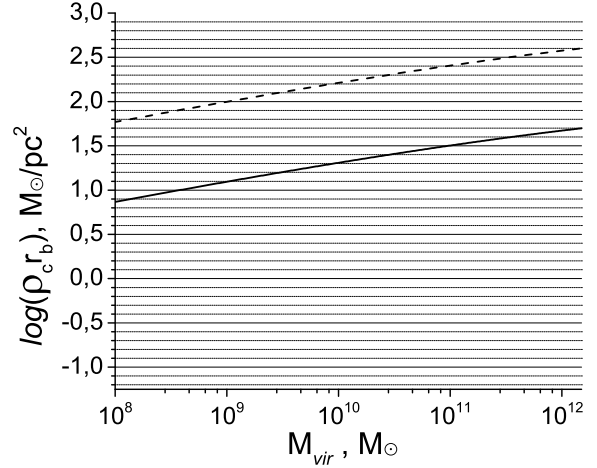}}
\caption{Value (\ref{16a26}) of $\rho_c r_b$ for various halo masses (solid line). For the entire
range of galactic masses ($M_{vir}\simeq 10^9 - 10^{12} M_\odot$) multiplication $\rho_c r_b$
remains almost constant. According to Eq.~(\ref{16a17}), $\rho_c r_b$ is inversely proportional to
$\langle\alpha\rangle$. The dashed line represents $\rho_c r_b$, if we use a lower value of
$\langle\alpha\rangle$ (Eq.~(\ref{16a27}) instead of (\ref{16a18})). This agrees well with
observations ($\log(\rho_c r_b)=2.15\pm 0.2$ in units of $\log (M_\odot \text{pc}^{-2})$).}
 \label{16fig3}
\end{figure}

It is convenient to introduce $\varrho$ --- the average halo density: $\mvir=\frac43 \pi \rvir^3
\varrho$. Substituting Eqs.~(\ref{16a18}) and (\ref{16a19}) to (\ref{16a17}), we obtain
 \begin{equation}
 \label{16a20}
\rho_c r_c = \frac{2^{16/3}}{\pi^{4/3} 3^{2/3}} \mvir^{1/3} {\varrho}^{2/3}\simeq 4.21 \mvir^{1/3}
{\varrho}^{2/3}.
\end{equation}
As $\mvir$ grows, multiplier $\mvir^{1/3}$ slowly increases, while ${\varrho}^{2/3}$ slowly
decreases, since smaller halos formed at higher $z$, when the Universe density was higher. Roughly
speaking, the density of a structure is proportional to the density of the Universe at the moment
$z$ when it collapsed \citep{halomodel}, that is, ${\varrho}\propto (z+1)^{3}$. Indeed, structures
form when their density contrast $\delta\rho/\rho$ reaches a certain value (close to $1$) that does
not depend on the mass \citep[section 7.2.2]{gorbrub2}. Strictly speaking, the dependence of $z$ on
$\mvir$ is ambiguous: the initial perturbations can be considered as a random Gaussian field, and
structures of the same mass can collapse at different $z$. Nevertheless, we may consider an
averaged redshift $z$ of the collapse of structures of mass $\mvir$.

We have accepted for the Milky Way $\mvir=M_{MW}=10^{12} M_\odot$, $\rvir=250$~{kpc}
\citep{klypin}. It corresponds to $\varrho_{MW}=1.5\cdot 10^4 M_\odot \text{kpc}^{-3}$. Instead of
$\mvir$, we can characterize the structures by the present-day wave number $k$ of the primordial
perturbations from which they were formed. Of course, $\mvir\propto k^{-3}$. The Milky Way mass
$\mvir\simeq 10^{12} M_\odot$ corresponds to $k_{MW}\simeq (0.6 \text{Mpc})^{-1}$ \citep{gorbrub2}.
So $k=k_{MW} (\mvir/M_{MW})^{-1/3}$.

The shape of dependence $z(\mvir)$ is defined by the cosmological model. We consider the very
standard $\Lambda$CDM scenario with $H_0=67.3 \text{(Mpc$^{-1}$ km/s)}$ (i.e., $H^{-1}_0=4.58\cdot
10^{17}$~s), $\Omega_m=0.315$ \citep{planck}. In this case, $z$ logarithmically depends on $\mvir$
\citep[eq. 5.47]{gorbrub2}
 \begin{equation}
 \label{16a21}
z+1\propto \ln\left(\dfrac{0.4(\sqrt2-1)c}{H_0 \sqrt{\Omega_m} \sqrt{z_{eq}+1}} k \right).
\end{equation}
We accept $z_{eq}=3100$ for the equi-density redshift of radiation and matter \citep{gorbrub2}.
Substituting here ${\varrho}\propto (z+1)^{3}$ and $k=k_{MW} (\mvir/M_{MW})^{-1/3}$, we obtain
 \begin{equation}
 \label{16a22}
\varrho\propto \ln^3\left[\dfrac{0.4(\sqrt2-1) c k_{MW}}{H_0 \sqrt{\Omega_m} \sqrt{z_{eq}+1}}
\left(\dfrac{\mvir}{M_{MW}}\right)^{-1/3}\right].
\end{equation}
It is convenient to introduce
 \begin{equation}
 \label{16a23}
X\equiv \dfrac{0.4(\sqrt2-1) c k_{MW}}{H_0 \sqrt{\Omega_m} \sqrt{z_{eq}+1}}.
\end{equation}
Then Eq.~(\ref{16a22}) may be rewritten as
 \begin{equation}
 \label{16a24}
\varrho\propto \ln^3\left[X \left(\mvir/M_{MW}\right)^{-1/3}\right]
\propto\left(1-\dfrac{\ln\left(\mvir/M_{MW}\right)}{3\ln X} \right)^3.
\end{equation}
Consequently,
 \begin{equation}
 \label{16a25}
\varrho=\varrho_{MW} \left(1-\dfrac{\ln\left(\mvir/M_{MW}\right)}{3\ln X} \right)^3.
\end{equation}
Now we can insert this value into Eq.~(\ref{16a20}). To compare the result with observations, we
need to calculate $\rho_c r_b$ instead of $\rho_c r_c$. By using a profile-independent definition
of the core radius (\ref{16a0}), we obtain $r_{core}=r_b/2$ and $r_{core}\simeq 0.924 r_c$.
Consequently, $\rho_c r_c\simeq 0.541 \rho_c r_b$, and we derive the final result
 \begin{equation}
 \label{16a26}
\rho_c r_b = 7.78 {\varrho}^{2/3}_{MW} M^{1/3}_{MW} \left(\dfrac{\mvir}{M_{MW}}\right)^{1/3}
\left(1-\dfrac{\ln\left(\mvir/M_{MW}\right)}{3\ln X} \right)^2.
\end{equation}
Observations suggest that $\log(\rho_c r_b)\simeq\text{\it const}=2.15\pm 0.2$ in units of $\log
(M_\odot \text{pc}^{-2})$ for a large array of elliptic and spiral galaxies, and probably for the
Local Group dwarf spheroidal galaxies \citep{salucci2009}. Fig.~\ref{16fig3} represents the
dependence (\ref{16a20}) (solid line) predicted by the moderate relaxation model. Clearly, the
multiplication $\rho_c r_b$ is not perfectly constant; however, it changes only by a factor of
three when the virial mass of galaxies varies from $10^{9} M_\odot$ to $10^{12} M_\odot$, which
covers almost the entire galaxy mass range. This means that the variation of $\rho_c r_b$ does not
exceed the $3\sigma$ interval of the observations, and from this point of view, $\rho_c r_b$ may be
considered as constant.

Thus, the moderate evolution model naturally predicts an almost constant multiplication $\rho_c
r_b$ in the galaxy mass range, which agrees well with observations. On the other hand, the value of
the constant $\rho_c r_b$ predicted by Eq.~(\ref{16a20}) is lower approximately by a factor of
seven than the observed value. The contradiction may be obviated if we assume that supposition
(\ref{16a18}) of the value of $\langle\alpha\rangle$ is not true. Indeed, equation (\ref{16a18})
implies that $\langle\alpha\rangle$ has the highest possible value. This is not necessarily so;
moreover, there are some reasons to believe that the mean-square-root angular momentum of the
particles of DM halos is fairly small \citep{11}. According to Eq.~(\ref{16a17}), $\rho_c r_b$ is
inversely proportional to $\langle\alpha\rangle$. If we insert
\begin{equation}
 \label{16a27}
 \langle\alpha\rangle=\frac{1}{28} \sqrt{G M_{vir} R_{vir}}
\end{equation}
instead of Eq.~(\ref{16a18}) into (\ref{16a17}), we obtain the dependence of $\rho_c r_b$ on
$\mvir$ in excellent agreement with observations (the dashed line in Fig.~\ref{16fig3}).

It is important to underline that concluding about the constancy of the multiplication $\rho_c r_b$
for the galaxy mass objects is an inherent property of the moderate-relaxation model and does not
depend on the choice of constants in Eqs.~(\ref{16a18}) and (\ref{16a27}). For more massive halos
($\mvir\ge 10^{13} M_\odot$) the model predicts an even weaker dependence of $\rho_c r_b$ on
$\mvir$; however, the model itself is hardly adequate for objects this massive. Very small halos
($\mvir< 10^{6} M_\odot$) should have $\rho_c r_b\propto \mvir^{1/3}$, that is, $\rho_c r_b$ is not
quite constant anymore. However, the dependence remains rather weak.

\section{Conclusion}
\label{16sec5} Thus assuming moderate relaxation in the formation of dark matter halos invariably
leads to density profiles that match those observed in the central regions of galaxies. The profile
is insensitive to the initial conditions. It has a central core; in the core area the density
distribution behaves as an Einasto profile with a low index ($n\sim 0.5$). At larger distances it
has an extended region with $\rho\propto r^{-2}$. The multiplication of the central density by the
core radius is almost independent of the halo mass.

This is exactly the shape that observations suggest for the central region of galaxies. On the
other hand, it does not fit the galaxy cluster profiles. This implies that the relaxation of huge
objects ($\mvir>10^{13} M_\odot$, galaxy clusters) is violent. However, it is moderate for galaxies
or smaller objects.

The most plausible explanation of this fact is that the concentrations $\cvir$ of small halos are
much higher. As we showed, for the relaxation to be violent, the energies of a significant part of
the particles need to change by a factor $\sim \cvir$ with respect to the initial values (so that
these particles have $r_0\simeq 0$ and form the cusp). Consequently, for galaxy clusters the
relaxation is violent if the energies of the particles decrease by a factor $3-5$ ($\cvir=3-5$ for
these objects). Such a change seems possible even in a collisionless system. However, $\cvir\sim
15$ even for massive galaxies and can be much higher for dwarf objects. Then the violent relaxation
claims that the energies of a significant fraction of the particles change by a factor $15-20$:
this requirement is quite strong. Probably, some other factors, such as the baryon component, also
influence the intensity of the relaxation. This question merits a more detailed consideration.
\begin{figure}
 \resizebox{\hsize}{!}{\includegraphics[angle=0]{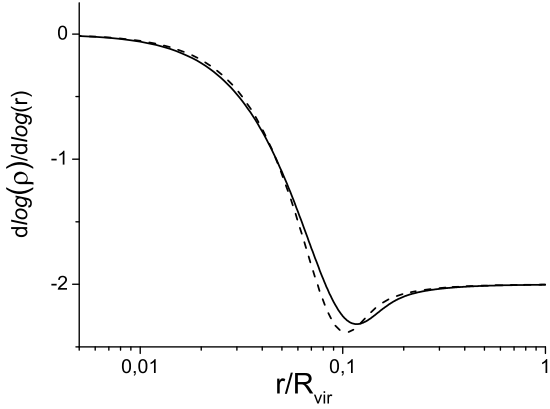}}
\caption{Shapes of density profiles calculated with the help of the exact equation (\ref{16a13})
(solid line) and approximation (\ref{16a15}) (dashed line, see the Appendix for details).}
 \label{16fig4}
\end{figure}

\begin{acknowledgements}
Financial support by Bundesministerium f\"ur Bildung und Forschung through DESY-PT, grant 05A11IPA,
is gratefully acknowledged. BMBF assumes no responsibility for the contents of this publication. We
acknowledge support by the Helmholtz Alliance for Astroparticle Physics HAP funded by the
Initiative and Networking Fund of the Helmholtz Association.
 \end{acknowledgements}

\appendix
\label{append16}
\section{A convolution of two functions: how does one transform equation (\ref{16a13}) into (\ref{16a15})?}
We consider a convolution of two functions $\int g_1(x) g_2(x,y) dx$ where $g_1(x)$ is finite, that
is, $g_1(x)$ differs noticeably from zero only in a narrow interval $x\in [x_1,x_2]$, and
$g_2(x,y)$ depend only slightly on $x$ at this interval for any given $y$. Then we may roughly
estimate
\begin{equation}
 \label{16a29}
\int g_1(x) g_2(x,y) dx\simeq g_2(\langle x\rangle,y) \int g_1(x)dx,
 \end{equation}
where $\langle x\rangle$ is the value of $x$ averaged over $[x_1,x_2]$. Indeed, equation
(\ref{16a29}) becomes exact if the width of $g_1(x)$ is negligible $\int \delta(x-x_0) g_2(x,y) dx
= g_2(x_0,y)$. It is also exact if $g_2(x,y)$ does not change at all, when $x$ runs between $x_1$
and $x_2$: $\int g_1(x) g_2(y) dx = g_2(y) \int g_1(x)dx$. If $g_2(x,y)$ depends on $x$ at
$[x_1,x_2]$, equation (\ref{16a29}) is an approximation. However, our prime interest here is the
accuracy of transformation (\ref{16a13}) into (\ref{16a15}) with the help of equation
(\ref{16a29}). We show below that the approximation is quite good in this case.

First of all, we should determine the best way to find $r_c$ (see Eq.~(\ref{16a14})), that is, to
average $\dfrac{\sqrt{2 |\phi(0)|}}{\alpha(r_0)}$ over the halo. Since we are interested in the
very central region of the halo, we consider the case when $r\to 0$. The Dawson function
$D(x)\simeq x$ if $x$ is small, and we obtain from Eq.~(\ref{16a13})
 \begin{equation}
 \label{16a30}
\rho_c = \frac{1}{2 \pi}\int^\infty_0\!\frac{f(r_0)}{\alpha(r_0) T(r_0)}\; \dfrac{\sqrt{2
|\phi(0)|}}{\alpha(r_0)} dr_0.
\end{equation}
Dividing this equation by (\ref{16a16}), we obtain
 \begin{equation}
 \label{16a31}
\frac{1}{r_c} = \frac{\int^\infty_0\!\dfrac{f(r_0)}{\alpha(r_0) T(r_0)}\; \dfrac{\sqrt{2
|\phi(0)|}}{\alpha(r_0)} dr_0}{\int^\infty_0\!\dfrac{f(r_0) dr_0}{\alpha(r_0) T(r_0)}}.
\end{equation}
This equation can be considered as a sort of averaging of function $\frac{\sqrt{2
|\phi(0)|}}{\alpha(r_0)}$ over the halo with the use of $\frac{f(r_0)}{\alpha(r_0) T(r_0)}$ as the
weighting function. This method of averaging yields the best fitting of Eq.~(\ref{16a13}) in the
halo center.

To estimate the accuracy of transformation (\ref{16a13}) into (\ref{16a15}), we may consider a more
or less realistic model of functions $f(r_0)$, $\alpha(r_0)$, $T(r_0)$, and then to compare the
density profiles obtained with the help of Eqs.~(\ref{16a13}) and (\ref{16a15}). We assume that
$f(r_0)$ has a Gaussian shape
 \begin{equation}
 \label{16a32}
\begin{cases}
f(r_0)\propto\exp\left(-\dfrac{(r_0-a)^2}{2\sigma^2}\right),&r_0>0.1 \rvir\\
f(r_0)=0,&r_0<0.1 \rvir,
\end{cases}
  \end{equation}
where $a=0.7\rvir$, $\sigma=0.2\rvir$. The parameters are chosen so that the total final energy of
the halo is approximately equal to the initial one, $f(r_0)$ is finite, as described at the end of
the {\it Energy distribution} section, and the region $\dfrac{2\rvir}{\cvir}$ falls approximately
outside the $3\sigma$ area, if $\cvir\sim 15$, which is a typical value for galaxies. $T(r_0)$ is
mainly defined by the particle motion near the apocenter \citep{14}. Since $M(r_0)\simeq\mvir$ if
$r_0\sim \rvir$, we may assume that $T(r_0)\propto\sqrt{r_0}$, as in the case of a point mass
$\mvir$. The same reasoning allows us to assume by analogy with Eqs.~(\ref{16a18}) and
(\ref{16a27}) that $\alpha(r_0)\propto \sqrt{M(r_0) r_0}\propto \sqrt{r_0}$, and that $\sqrt{2
(\phi(r_0)-\phi(0))}\simeq \sqrt{2 |\phi(0)|}$. A normalization of $f(r_0)$, $\alpha(r_0)$, and
$T(r_0)$ is not significant, since we are interested in the profile shape, and we plot it in
$\log\rho/\log r$ coordinates. However, we choose a proper value of $\sqrt{2 |\phi(0)|}$ to obtain
a desirable value of $r_c$ according to equation (\ref{16a31}). We chose $r_c=0.05\rvir$, exactly
as in Fig.~\ref{16fig2}.

Fig.~\ref{16fig4} represents the density profiles calculated for this model with the help of exact
equation (\ref{16a13}) (solid line) and approximation (\ref{16a15}) (dashed line). Clearly, the
deviations are quite small, especially in the halo center and at large radii. Moreover,
approximation (\ref{16a15}) has the same shape as the exact solution, that is, the same
Einasto-like profile in the center, the same core radius, and the same $\rho\propto r^{-2}$ region.
Consequently, the conclusions of this paper are valid for the exact solution (\ref{16a13}) as well.
Thus the approximation of equation (\ref{16a13}) by (\ref{16a15}) is quite accurate for realistic
models of halo parameters.


\begin{thebibliography}{44}
\expandafter\ifx\csname natexlab\endcsname\relax\def\natexlab#1{#1}\fi

\bibitem[{{Baushev}(2011)}]{11}
{Baushev}, A.~N. 2011, \mnras, 417, L83

\bibitem[{{Baushev}(2012)}]{clump}
{Baushev}, A.~N. 2012, \mnras, 420, 590

\bibitem[{{Baushev}(2013{\natexlab{a}})}]{14}
{Baushev}, A.~N. 2013{\natexlab{a}}, \apj, 771, 117

\bibitem[{{Baushev}(2013{\natexlab{b}})}]{13}
{Baushev}, A.~N. 2013{\natexlab{b}}, ArXiv:1312.0314

\bibitem[{{Baushev}(2014)}]{15}
{Baushev}, A.~N. 2014, \apj, 786, 65

\bibitem[{{Binney} \& {Tremaine}(2008)}]{bt}
{Binney}, J. \& {Tremaine}, S. 2008, {Galactic Dynamics: Second Edition}
  (Princeton University Press)

\bibitem[{{Burkert}(1995)}]{burkert}
{Burkert}, A. 1995, \apjl, 447, L25

\bibitem[{{Carney} \& {Latham}(1987)}]{escape}
{Carney}, B.~W. \& {Latham}, D.~W. 1987, in IAU Symposium, Vol. 117, Dark
  matter in the Universe, ed. J.~{Kormendy} \& G.~R. {Knapp}, 39--48

\bibitem[{{Chemin} {et~al.}(2011){Chemin}, {de Blok}, \& {Mamon}}]{mamon2011}
{Chemin}, L., {de Blok}, W.~J.~G., \& {Mamon}, G.~A. 2011, \aj, 142, 109

\bibitem[{{Cooray} \& {Sheth}(2002)}]{halomodel}
{Cooray}, A. \& {Sheth}, R. 2002, \physrep, 372, 1

\bibitem[{{de Blok} \& {Bosma}(2002)}]{bosma2002}
{de Blok}, W.~J.~G. \& {Bosma}, A. 2002, \aap, 385, 816

\bibitem[{{de Blok} {et~al.}(2001){de Blok}, {McGaugh}, \&
  {Rubin}}]{deblok2001}
{de Blok}, W.~J.~G., {McGaugh}, S.~S., \& {Rubin}, V.~C. 2001, \aj, 122, 2396

\bibitem[{{Diemand} \& {Kuhlen}(2008)}]{diemandkuhlen2008}
{Diemand}, J. \& {Kuhlen}, M. 2008, \apjl, 680, L25

\bibitem[{{Diemand} {et~al.}(2007){Diemand}, {Kuhlen}, \&
  {Madau}}]{diemand2007}
{Diemand}, J., {Kuhlen}, M., \& {Madau}, P. 2007, \apj, 667, 859

\bibitem[{{Diemand} {et~al.}(2005){Diemand}, {Madau}, \& {Moore}}]{diemand2005}
{Diemand}, J., {Madau}, P., \& {Moore}, B. 2005, \mnras, 364, 367

\bibitem[{{Donato} {et~al.}(2009){Donato}, {Gentile}, {Salucci}, {Frigerio
  Martins}, {Wilkinson}, {Gilmore}, {Grebel}, {Koch}, \& {Wyse}}]{salucci2009}
{Donato}, F., {Gentile}, G., {Salucci}, P., {et~al.} 2009, \mnras, 397, 1169

\bibitem[{{Evans} \& {Collett}(1997)}]{evans1997}
{Evans}, N.~W. \& {Collett}, J.~L. 1997, \apjl, 480, L103

\bibitem[{{Gentile} {et~al.}(2007){Gentile}, {Salucci}, {Klein}, \&
  {Granato}}]{gentile2007}
{Gentile}, G., {Salucci}, P., {Klein}, U., \& {Granato}, G.~L. 2007, \mnras,
  375, 199

\bibitem[{{Gorbunov} \& {Rubakov}(2010)}]{gorbrub2}
{Gorbunov}, D.~S. \& {Rubakov}, V.~A. 2010, Introduction to the Early Universe
  theory. Volume 2: Cosmological perturbations. (LKI publishing house, Moscow)

\bibitem[{{Governato} {et~al.}(2012){Governato}, {Zolotov}, {Pontzen},
  {Christensen}, {Oh}, {Brooks}, {Quinn}, {Shen}, \& {Wadsley}}]{governato2012}
{Governato}, F., {Zolotov}, A., {Pontzen}, A., {et~al.} 2012, \mnras, 422, 1231

\bibitem[{{Hansen} {et~al.}(2006){Hansen}, {Moore}, {Zemp}, \&
  {Stadel}}]{2006JCAP...01..014H}
{Hansen}, S.~H., {Moore}, B., {Zemp}, M., \& {Stadel}, J. 2006, \jcap, 1, 14

\bibitem[{{Hayashi} {et~al.}(2003){Hayashi}, {Navarro}, {Taylor}, {Stadel}, \&
  {Quinn}}]{hayashi2003}
{Hayashi}, E., {Navarro}, J.~F., {Taylor}, J.~E., {Stadel}, J., \& {Quinn}, T.
  2003, \apj, 584, 541

\bibitem[{{Klypin} {et~al.}(2013){Klypin}, {Prada}, {Yepes}, {Hess}, \&
  {Gottlober}}]{klypin2013}
{Klypin}, A., {Prada}, F., {Yepes}, G., {Hess}, S., \& {Gottlober}, S. 2013,
  ArXiv e-prints

\bibitem[{{Klypin} {et~al.}(2002){Klypin}, {Zhao}, \& {Somerville}}]{klypin}
{Klypin}, A., {Zhao}, H., \& {Somerville}, R.~S. 2002, \apj, 573, 597

\bibitem[{{Kormendy} \& {Freeman}(2004)}]{kormendy2004}
{Kormendy}, J. \& {Freeman}, K.~C. 2004, in IAU Symposium, Vol. 220, Dark
  Matter in Galaxies, ed. S.~{Ryder}, D.~{Pisano}, M.~{Walker}, \&
  K.~{Freeman}, 377

\bibitem[{{Kuhlen} {et~al.}(2010){Kuhlen}, {Weiner}, {Diemand}, {Madau},
  {Moore}, {Potter}, {Stadel}, \& {Zemp}}]{kuhlen2010}
{Kuhlen}, M., {Weiner}, N., {Diemand}, J., {et~al.} 2010, \jcap, 2, 30

\bibitem[{{Landau} \& {Lifshitz}(1980)}]{ll10}
{Landau}, L.~D. \& {Lifshitz}, E.~M. 1980, {Statistical physics. Pt.1, Pt.2}

\bibitem[{{Lynden-Bell}(1967)}]{violent}
{Lynden-Bell}, D. 1967, \mnras, 136, 101

\bibitem[{{Marchesini} {et~al.}(2002){Marchesini}, {D'Onghia}, {Chincarini},
  {Firmani}, {Conconi}, {Molinari}, \& {Zacchei}}]{marchesini2002}
{Marchesini}, D., {D'Onghia}, E., {Chincarini}, G., {et~al.} 2002, \apj, 575,
  801

\bibitem[{{Marochnik} \& {Suchkov}(1984)}]{suchkov}
{Marochnik}, L.~S. \& {Suchkov}, A.~A. 1984, {The Galaxy}

\bibitem[{{Moore} {et~al.}(1999){Moore}, {Quinn}, {Governato}, {Stadel}, \&
  {Lake}}]{moore1999}
{Moore}, B., {Quinn}, T., {Governato}, F., {Stadel}, J., \& {Lake}, G. 1999,
  \mnras, 310, 1147

\bibitem[{{Navarro} {et~al.}(2010){Navarro}, {Ludlow}, {Springel}, {Wang},
  {Vogelsberger}, {White}, {Jenkins}, {Frenk}, \& {Helmi}}]{navarro2010}
{Navarro}, J.~F., {Ludlow}, A., {Springel}, V., {et~al.} 2010, \mnras, 402, 21

\bibitem[{{Neto} {et~al.}(2007){Neto}, {Gao}, {Bett}, {Cole}, {Navarro},
  {Frenk}, {White}, {Springel}, \& {Jenkins}}]{neto2007}
{Neto}, A.~F., {Gao}, L., {Bett}, P., {et~al.} 2007, \mnras, 381, 1450

\bibitem[{{Oh} {et~al.}(2011){Oh}, {de Blok}, {Brinks}, {Walter}, \&
  {Kennicutt}}]{oh2011}
{Oh}, S.-H., {de Blok}, W.~J.~G., {Brinks}, E., {Walter}, F., \& {Kennicutt},
  Jr., R.~C. 2011, \aj, 141, 193

\bibitem[{{Okabe} {et~al.}(2010){Okabe}, {Zhang}, {Finoguenov}, {Takada},
  {Smith}, {Umetsu}, \& {Futamase}}]{okabe2010}
{Okabe}, N., {Zhang}, Y.-Y., {Finoguenov}, A., {et~al.} 2010, \apj, 721, 875

\bibitem[{{Planck Collaboration} {et~al.}(2013){Planck Collaboration}, {Ade},
  {Aghanim}, {Armitage-Caplan}, {Arnaud}, {Ashdown}, {Atrio-Barandela},
  {Aumont}, {Baccigalupi}, {Banday}, \& et~al.}]{planck}
{Planck Collaboration}, {Ade}, P.~A.~R., {Aghanim}, N., {et~al.} 2013, ArXiv
  e-prints

\bibitem[{{Power} {et~al.}(2003){Power}, {Navarro}, {Jenkins}, {Frenk},
  {White}, {Springel}, {Stadel}, \& {Quinn}}]{power2003}
{Power}, C., {Navarro}, J.~F., {Jenkins}, A., {et~al.} 2003, \mnras, 338, 14

\bibitem[{{Quinlan}(1996)}]{quinlan1996}
{Quinlan}, G.~D. 1996, \na, 1, 255

\bibitem[{{Rubin} {et~al.}(1978){Rubin}, {Thonnard}, \& {Ford}}]{rubinold}
{Rubin}, V.~C., {Thonnard}, N., \& {Ford}, Jr., W.~K. 1978, \apjl, 225, L107

\bibitem[{{Salucci} {et~al.}(2007){Salucci}, {Lapi}, {Tonini}, {Gentile},
  {Yegorova}, \& {Klein}}]{salucci2007}
{Salucci}, P., {Lapi}, A., {Tonini}, C., {et~al.} 2007, \mnras, 378, 41

\bibitem[{{Sofue} \& {Rubin}(2001)}]{rubinnew}
{Sofue}, Y. \& {Rubin}, V. 2001, \araa, 39, 137

\bibitem[{{Stadel} {et~al.}(2009){Stadel}, {Potter}, {Moore}, {Diemand},
  {Madau}, {Zemp}, {Kuhlen}, \& {Quilis}}]{mo09}
{Stadel}, J., {Potter}, D., {Moore}, B., {et~al.} 2009, \mnras, 398, L21

\bibitem[{{Tollerud} {et~al.}(2012){Tollerud}, {Beaton}, {Geha}, {Bullock},
  {Guhathakurta}, {Kalirai}, {Majewski}, {Kirby}, {Gilbert}, {Yniguez},
  {Patterson}, {Ostheimer}, {Cooke}, {Dorman}, {Choudhury}, \&
  {Cooper}}]{tollerud2012}
{Tollerud}, E.~J., {Beaton}, R.~L., {Geha}, M.~C., {et~al.} 2012, \apj, 752, 45

\bibitem[{{Wang} {et~al.}(2012){Wang}, {Frenk}, {Navarro}, {Gao}, \&
  {Sawala}}]{wang2012}
{Wang}, J., {Frenk}, C.~S., {Navarro}, J.~F., {Gao}, L., \& {Sawala}, T. 2012,
  \mnras, 424, 2715

\end{thebibliography}

\end{document}